\documentclass[11pt,a4paper]{article}
\usepackage{jcappub}
\usepackage{bm}
\usepackage{amsmath}
\usepackage{amssymb}
\usepackage{graphicx}
\usepackage{hyperref}
\usepackage{color}
\usepackage{mathtools}
\usepackage{times}
\usepackage{hyperref}
\usepackage{comment}
\hypersetup{
    colorlinks=true,
    linkcolor=blue,
    filecolor=magenta,
    citecolor=blue,
    urlcolor=blue,
}

\newcommand{\bdv}[1]{{\bf #1}}
\newcommand{\RA}{\rightarrow}
\newcommand{\mpc}{{\rm Mpc}}
\newcommand{\up}[1]{{\rm #1}}
\newcommand{\dgobs}{\delta_g^\up{obs}}
\newcommand{\ngobs}{n_g^\up{obs}}
\newcommand{\ngavg}{\bar n_{g,\Omega}^\up{obs}}
\newcommand{\den}{\delta n}
\newcommand{\MMM}{{\cal M}}
\newcommand{\beeq}{\begin{equation}}
\newcommand{\eneq}{\end{equation}}
\newcommand{\bear}{\begin{eqnarray}}
\newcommand{\enar}{\end{eqnarray}}
\newcommand{\para}{\parallel}
\newcommand{\Dquad}{\qquad\qquad}
\newcommand{\Tquad}{\qquad\qquad\qquad}
\newcommand{\hmpc}{{h^{-1}\mpc}}
\newcommand{\OO}{\mathcal{O}}
\newcommand{\kvec}{\bdv{k}}
\newcommand{\svec}{\bdv{s}}
\newcommand{\xvec}{\bdv{x}}
\newcommand{\Vang}{\bdv{\hat n}}
\newcommand{\Nang}{\hat{\bm{n}}}
\newcommand{\al}{\alpha}
\newcommand{\ga}{\gamma}
\newcommand{\de}{\delta}

\begin{document}

\begin{titlepage}

\setcounter{page}{1} \baselineskip=15.5pt \thispagestyle{empty}
\pagenumbering{roman}

\bigskip

\vspace{1cm}
\begin{center}
{\fontsize{20}{28}\selectfont \bfseries
Robustness of the Baryonic Acoustic Oscillation Peak Position
against Monopole Fluctuations in Galaxy Surveys}
\end{center}

\vspace{0.2cm}

\begin{center}
{\fontsize{13}{30}\selectfont 
Christopher Magnoli$^a$, Giovanni Piccoli$^a$, Jaiyul Yoo$^{a,b}$,
Daniel Eisenstein$^c$}
\end{center}

\begin{center}
\vskip 8pt
\textsl{$^a$ Center for Theoretical Astrophysics and Cosmology,
Department of Astrophysics}\\
\textsl{University of Z\"urich, Winterthurerstrasse 190,
CH-8057, Z\"urich, Switzerland}

\vskip 7pt

\textsl{$^b$Department of Physics, University of Z\"urich,
Winterthurerstrasse 190, CH-8057, Z\"urich, Switzerland}

\vskip 7pt

\textsl{$^c$Harvard-Smithsonian Center for Astrophysics, 
60 Garden Street, Cambridge, MA 02138, USA}

\vskip 7pt

\today

\end{center}

\note{christopher.magnoli@uzh.ch}
\note{giovanni.piccoli@uzh.ch}
\note{jaiyul.yoo@uzh.ch}
\note{deisentstein@cfa.harvard.edu}

\vspace{1.2cm}
\hrule \vspace{0.3cm}
\noindent {\sffamily \bfseries Abstract} \\[0.1cm]
Precise measurements of the baryonic acoustic oscillation (BAO) peak position
are one of the key goals in modern galaxy surveys. We investigate the impact
of the monopole fluctuations in galaxy surveys on the BAO peak position
in the galaxy
 two-point redshift-space  multipole-correlation functions. The background
mean density of a galaxy sample is estimated from the survey itself by
averaging the observed galaxy number density, and this averaged number density
(monopole)
is subtracted to obtain the observed galaxy number density fluctuation,
which includes an additional contribution from the monopole fluctuations,
propagating to galaxy clustering analysis.
We compute the contributions of the monopole
fluctuations to the redshift-space multipole-correlation functions.
While the monopole contributions in galaxy two-point correlation function
are anisotropic and non-negligible  in amplitude, the contributions are 
so smooth around the BAO peak position that they are almost completely
absorbed in the BAO broadband model, leaving the BAO peak position unaffected
at the sub-percent level. Our results demonstrate the robustness of the
BAO analysis against the unknown systematic error.
\vskip 10pt
\hrule

\vspace{0.6cm}
\end{titlepage}

\clearpage

\noindent\hrulefill \tableofcontents \noindent\hrulefill

\pagenumbering{arabic}

\section{Introduction}
Measurements of the Baryon Acoustic Oscillation (BAO) signals in galaxy
clustering are one of the principal methods to map the expansion history 
of the Universe in modern galaxy surveys (see, e.g.,
\cite{EIHU98,EISEWH07,WEMOET13} for recent reviews).
Originating from the tight coupling of photons and baryons in the early 
Universe \cite{PEYU70,SUZE70a}, these sound waves in the baryon-photon plasma
leave a characteristic imprint in both the cosmic microwave background (CMB) 
anisotropies and galaxy clustering.
The BAO signature in galaxy clustering manifests itself
 as a distinct peak in the galaxy two-point correlation function or
 a series of oscillations in the galaxy power spectrum. 
Since the comoving scale of the BAO signature is determined 
by the sound horizon 
at the end of the baryon drag epoch and is precisely constrained 
by CMB observations \cite{PLANCKcos18,PLANCKover18},
measurements of the BAO signature in the observed
redshift and angle provide robust measurements of the Hubble parameter
and angular diameter distance to the redshift of the galaxy sample.
Using the measurements from the Sloan Digital Sky Survey (SDSS) \cite{YOADET00},
the first BAO detection was achieved in 2005 \cite{EIZEET05}.
In the following years, observational cosmology has seen 
a dramatic increase in the development of large-scale galaxy surveys
with spectroscopic measurements and increasingly larger redshift ranges,
such as the 6-degree Field Galaxy Survey (6dFGS) \cite{BEBLET11},
WiggleZ \cite{BLBRET11}, and the Dark Energy Spectroscopic Instrument (DESI) 
\cite{DESI13}.
The success of these programs  enabled precise measurements of the distance
scales \cite{EIZEET05,BLCOET07,PEREET10,
PEREET10,ANAUET12,ATBAET18,BAPAET21,DESIbao24,DESIcos24}.
In the coming years, the Dark Energy Spectroscopic Instrument (DESI) 
and its extension will
achieve the sub-percent level precision measurements of the distance scales.

The level of measurement precision in the current and upcoming surveys
demands that our theoretical modeling of galaxy clustering need to be made 
accurate at the same level of measurement 
precision to avoid any misinterpretation of the 
measurements from any systematic errors in the theoretical modeling.
In recent work \cite{YOEI25}, it was shown that there exists a new unknown
contribution, called the monopole fluctuations, 
to the observed galaxy clustering that can affect the BAO measurements. 
Since the background mean number density of any galaxy samples is unknown,
the galaxy survey itself needs to estimate the mean number density for the
subsequent clustering analysis, and this estimate obtained by averaging
over the angle at a given redshift includes the galaxy monopole fluctuation,
which then propagates to the observed galaxy number density
fluctuation and its two-point
correlation function. In particular, the monopole correction in the
observed galaxy two-point correlation function amounts to be a 
{\it few percent}
in amplitude of the correlation function at the BAO scale, and it affects
the correlation function differently along the line-of-sight and the transverse
directions \cite{YOEI25}.

Following-up the initial investigation in \cite{YOEI25}, we first
compute the observed redshift-space multipole correlation functions
and the contributions of the monopole fluctuations to the multipole
correlation functions. We then quantify the  shift of the BAO peak
position due to the monopole fluctuations by using the standard BAO fitting
procedure \cite{ROBEET17}
to remove the broadband power and any other systematic effects
and by accounting for the proper covariance matrix in a given galaxy survey.
The paper is organized as follows: Section~2.1 presents our analytical model
for the observed galaxy number density and the covariance matrix 
in galaxy surveys. In Sec.~2.2, 
we introduce the monopole fluctuations, following
the work in \cite{YOEI25}. We then briefly review the standard
procedure to estimate the covariance matrix and to
measure the BAO peak position in galaxy surveys in the following sections.
Section~3 presents our main findings of the impact of the monopole fluctuations
on the BAO peak position measurements. We first present the redshift-space
multipole correlation functions and the multipole contributions of the monopole
correction. Second, we apply the BAO fitting procedure to isolate the BAO
peak position from the broadband power by 
marginalizing over nuisance parameters.
Then, we quantify the shift of the BAO peak position due to the monopole 
fluctuations in the redshift-space galaxy monopole and quadrupole correlation
functions. Finally, the implications of our results are 
summarized and discussed in Sec.~4.

\section{Analytical Modeling of the Observed Galaxy Two-Point Correlation
and the BAO Analysis}
\label{sec:basics}
We describe the theoretical framework for our analytical
modeling of the observed galaxy two-point correlation function accounting
for the monopole fluctuations, and we briefly review
the standard BAO analysis method to isolate the BAO peak position and the
broad-band power in the galaxy two-point correlation function.

\subsection{Redshift-Space Galaxy Two-Point Correlation Function}
\label{sec:corr}
The galaxy two-point correlation function $\xi_g(\svec)$ characterizes how 
much galaxies cluster in space, and it represents the excess probability of 
finding galaxies at a given separation $\svec$, compared to the number of
galaxies expected from the background mean (see e.g., 
\cite{PEEBL80,PEACO99,DODEL03}).
Given the galaxy number density $n_g(\xvec,z)$ and its background mean 
$\bar{n}_g(z)$, the galaxy number density fluctuation $\delta_g(\mathbf{x})$ 
at redshift~$z$ can be constructed as
\beeq
\label{deltag}
\delta_g(\xvec,z) := \frac{n_g(\xvec,z) - \bar{n}_g(z)}{\bar{n}_g(z)}\,,
\eneq
and the galaxy two-point correlation function is defined 
in terms of $\delta_g(\xvec)$ as the ensemble average of galaxies at two
points separated by~$\svec$:
\beeq
\xi_g (\svec) = \langle \delta_g (\mathbf{x}) \delta_g (\mathbf{x+s})\rangle \, .
\eneq
The same clustering information can be expressed in Fourier space
by the galaxy power spectrum $P_g(\kvec)$
\beeq
\langle {\delta}_g(\kvec) {\delta}_g(\kvec')\rangle 
= (2\pi)^3 \delta^D (\kvec+\kvec') P_g (\kvec)\, ,
\eneq
where $\delta^D$ is the Dirac delta function. The Fourier mode~$\delta_g(\kvec)$
and the power spectrum~$P_g(\kvec)$ are the Fourier transform 
of~$\delta_g(\xvec)$ and~$\xi_g(\svec)$, respectively.
In this work, we model the observed galaxy number density 
fluctuation~$\de_g(\xvec,z)$ in redshift-space 
by using the Kaiser formula \cite{KAISE87}
\beeq
\label{KAISER}
\delta_g (\kvec,z) = \bigg[b(z) + f(z) \mu_k^2\bigg]\delta_m(\kvec,z)~,
\eneq
where $b(z)$ is the linear bias factor \cite{KAISE84},
$f(z)$~is the logarithmic growth rate, $\delta_m(\kvec,z)$ is the matter 
density fluctuation, and $\mu_k = \hat k\cdot \hat x$ is
the cosine angle between the line-of-sight 
direction~$\hat x$ and the wavevector~$\hat k$.
Here we adopt the linear theory  that the galaxy density fluctuation 
follows the matter density fluctuation on large scales \cite{KAISE84,BBKS86},
and
the redshift-space distortion arises due to the volume change  \cite{YOO09}
in redshift-space, compared to the real-space, while the relativistic
effects \cite{YOFIZA09, YOO14a} are ignored.

 Furthermore, we have adopted the distant-observer approximation, and 
the line-of-sight direction is unique for the galaxy sample, much larger than
any separation~$\svec$ of two galaxies. Under this approximation, 
the redshift-space power spectrum can be expanded in terms of 
Legendre polynomials $\mathcal{L}_l(\mu_k)$ into the redshift-space
multipole power spectra~$P_l(s)$ as
\beeq
\label{RSDDECOM}
P_g(k,\mu_k) = \left(b+f\mu_k^2\right)^2 P_m(k) = 
\sum_{l=0,2,4} \mathcal{L}_l(\mu_k) P_l(s)\,,
\eneq
where all the quantities are evaluated at the same mean
redshift $z_1\approx z_2\approx \bar{z}$ of galaxies.
Being the Fourier transform of the power spectrum, the galaxy two-point 
correlation function in redshift-space is then
\beeq
\label{ximul}
\xi_g(s, \mu) = \int \frac{d^3k}{(2\pi)^3}  \, P_g(k,\mu_k) e^{i \kvec\cdot\svec} 
= \sum_{l=0,2,4} \mathcal{L}_l(\mu) \xi_l(s)\,,
\eneq
where  $\mu = \hat s\cdot\hat x$ is the cosine angle between the line-of-sight
direction~$\hat x$ and the separation direction~$\hat s$. In the same way
the redshift-space power spectrum is decomposed, the redshift-space correlation
function~$\xi_g(s,\mu)$
was  decomposed \cite{HAMIL92,COFIWE94} in Equation~\eqref{ximul}
into the redshift-space multipole correlation~$\xi_l(s)$,
and by using the Kaiser formula, we obtain the redshift-space
galaxy multipole correlation functions 
(see \cite{HAMAET03,RESAET12,BESEET17} for recent measurements) as
\bear
\xi_0(s) &=& D^2\left[ b^2+\frac{2}{3} b \cdot f + \frac{1}{5} f^2\right] 
\int {d \ln{k}}\; \Delta^2_m(k) ~j_0(k s)\,,  \label{th monpole}\\
    \xi_2(s) &=&  -4 D^2\left[\frac{1}{3}b \cdot f+\frac{1}{7} f^2\right] 
\int {d \ln{k}}\; \Delta^2_m(k) ~ j_2(k s)\,, \label{th quad}\\
    \xi_4(s) &=& \frac{8}{35} D^2f^2 \int {d \ln{k}}\; \Delta^2_m(k) ~
 j_4(k s)\,,\label{th hexa}
\enar
where $j_l(x)$ is the spherical Bessel function, 
$\Delta_m^2(k)=k^3P_m(k)/2\pi^2$ is the dimensionless matter power spectrum
at~$z=0$, and $D$~is the growth factor normalized to unity at~$z=0$.

\subsection{Monopole Fluctuations in Galaxy Surveys}
\label{sec:mono}
We briefly describe the monopole fluctuations found in \cite{YOEI25}
and review the effect of the monopole fluctuations on the two-point correlation
function. In galaxy surveys, the galaxy number density $n_g^\up{obs}(z,\Vang)$
is measured by counting the number of galaxies in a small volume described 
by the observed redshift bin~$dz$ and the solid angle~$d\Omega$, and the
observed galaxy number density fluctuation
can then be obtained by using Eq.~\eqref{deltag},
provided that the background mean number density~$\bar n_g(z)$ is known.
In contrast to the background matter density $\bar{\rho}_m(z)$,
where its redshift evolution $\bar\rho_m\propto(1+z)^{3}$ is known and 
its amplitude today is a cosmological parameter,
the background galaxy mean number density $\bar n_g(z)$ 
in any survey is a priori unknown in terms of its amplitude at any time 
and the redshift evolution. In fact, galaxy surveys attempt to construct
galaxy samples as homogeneous over the redshift range as possible 
in their selection of the samples to minimize the complications from galaxy
formation physics and the subsequent evolution (see
\cite{EIANET01,COEIET08,WHBLET11,DESIbao24} for the observed mean number
density in galaxy surveys).

Hence, the observers estimate the background mean number density~$\bar n_g(z)$ 
from the survey itself by averaging the observed galaxy number density 
over the sky as
\beeq
\label{NBOBSOM}
\bar{n}_{g,\Omega}^\up{obs}(z) := \int \frac{d\Omega}{4 \pi}\; n_g^\text{obs}
(\Vang,z) = \bar{n}_g(z) \bigg[1 + \delta n(z)\bigg]\,,
\eneq
where the dimensionless monopole fluctuation~$\delta n (z)$ is defined 
as the angle-average of the galaxy number density fluctuation 
$\delta_g(\Vang,z)$  as
\beeq
\label{angmon}
\delta n(z) := \int \frac{d \Omega}{4 \pi}\; \delta_g(\Vang,z)\,,
\eneq
and we have assumed that the sky coverage of the surveys is full-sky.
Therefore, with our best estimate for the background mean number density 
$\bar{n}_{g,\Omega}^\text{obs}(z)$,
 one can rearrange the observed galaxy number density as
\beeq
n_g^\text{obs} (\Vang,z) =
 \bar{n}_{g,\Omega}^\text{obs}(z) \cdot \frac{1 +
 \delta_g(\mathbf{\hat{n}},z)}{1 + \delta n(z)}=:\bar n_{g,\Omega}^\up{obs}(z)
\bigg[1+\de_g^\up{obs}(\Vang,z)\bigg]~,
\eneq
where we defined the observed galaxy fluctuation we can obtain 
in any galaxy survey as
\beeq
\delta_g^\text{obs}(\Vang,z) := \frac{n_g^\text{obs}(\Vang,z)}
{\bar{n}_{g,\Omega}^\text{obs}(z)} - 1 = \frac{\delta_g(\Vang,z) - 
\delta n(z)}{1 + \delta n(z)} \simeq 
\delta_g(\Vang,z) - \delta n(z)+\OO(2)\,.
\eneq
This relation shows that the observed galaxy fluctuation is not just 
the standard galaxy fluctuation $\delta_g(\Vang,z)$ in redshift-space,
but includes the monopole fluctuation~$\delta n$, arising from the way 
the observers estimate the galaxy mean number density \cite{YOEI25}.
The monopole fluctuations inevitably exist in any galaxy surveys
as a form of the cosmic variance, because we can only perform angle average
of the fluctuation, rather than its full Euclidean average including
translation of the observer position (see \cite{MIYOET20} for details).

We finish this subsection by computing the observed two-point galaxy
correlation function, including the monopole fluctuations.
First, we decompose the galaxy number density fluctuation in terms of spherical
Harmonics~$Y_{l m}(\Vang)$ as
\beeq
\delta_g (\Vang,z) = \sum_{l m} a_{l m}(z) Y_{l m}(\Vang)~,\Tquad
a_{l m}(z) = \int d\Omega \, Y^*_{l m}(\Vang) \delta_g(\Vang,z)\, ,
\eneq
and the angular power spectrum $C_l$ is then defined as
\beeq
\langle a_{l m}(z_1)~a^*_{l' m'}(z_2) \rangle =\delta_{ll'}\delta_{mm'} 
C_l(z_1,z_2) ~.
\eneq
The angle average in Eq.~\eqref{angmon} for the monopole fluctuation 
$\delta n(z)$ isolates only the $l=0$ contribution (hence the name) as
\beeq
    \delta n (z) = \frac{1}{\sqrt{4\pi}}\, a_{00}(z)\,,\Tquad
\langle \delta n(z_1) \delta n(z_2)\rangle = \frac{1}{4\pi} C_0(z_1,z_2)\,.
\eneq
In the same way, we can compute the cross-correlation terms of~$\delta_g$ 
and~$\delta n$, and the observed galaxy two-point correlation function is 
then obtained \cite{YOEI25} as
\beeq
\label{monocorr}
\xi_g^{\text{obs}}(\svec,z_1,z_2)
=\xi_g(\svec,z_1,z_2) - \frac{1}{4\pi} C_0(z_1,z_2)~.
\eneq
Note that the first term is the standard galaxy two-point correlation function 
in redshift-space and the second term is the extra monopole correction.

\subsection{Baryon Acoustic Oscillation Analysis in the Two-Point Correlation
Function}
\label{sec:BAO}

Baryon Acoustic Oscillations (BAO) originate from pressure waves 
of the baryon-photon fluid in the early universe, and they
create a shell of excess baryons around the overdense region of dark matter.
The BAO scale~$r_d$ is the distance propagated by the acoustic
waves at the sound speed~$c_s$ over the age of the Universe until the
baryons decouple from the CMB photons at the drag epoch $z_d=1059$ 
\cite{EIHU98,EIHUET98,EIHU99,EIWH04}
\beeq
r_d= \int_{z_d}^{\infty}dz~ \frac{c_s(z)}{H(z)}~,\Tquad c_s^2={1\over3(1+R)}~,
\Tquad R:={3\bar\rho_b\over4\bar\rho_\ga}~,
\eneq
where $\bar\rho_b$ and $\bar\rho_\ga$ are the energy densities of the baryons
and photons.
This BAO scale is the standard ruler, well calibrated by the CMB observations
to 0.03\% level precision \cite{PLANCKcos18,PLANCKover18}. Since the BAO scale in
the observed galaxy two-point correlation function is identified 
in terms of the angular scale~$\Delta \theta$ and the redshift 
separation~$\Delta z$, their translation into the comoving separation into
the BAO scale~$r_d$ depends on the cosmological parameters adopted in the
fiducial model. With incorrect cosmological parameters in the fiducial
model, measurements of the BAO scale in the observed 
galaxy two-point correlation function
could result in a distortion from otherwise an isotropic ring of the BAO scale,
known as the the Alcock–Paczyński effect \cite{ALPA79}, characterized
in terms of two dilation parameters:
\beeq
    \alpha_\perp:=\frac{D_A(z)\,r_d^{\mathrm{fid}}}{D_A^{\mathrm{fid}}(z)\,r_d}~,
    \Tquad
    \alpha_\parallel:=\frac{H^{\mathrm{fid}}(z)\,r_d^{\mathrm{fid}}}{H(z)\,r_d}~,
\eneq
where $H(z)$ is the Hubble parameter and $D_A(z)$ is the angular diameter
distance.

Measuring the BAO peak position from the observed galaxy two-point correlation
function requires a template fitting procedure to isolate the BAO wiggle
from the broad-band power in the correlation function. We take
the standard procedure \cite{ROBEET17} adopted in the SDSS-III survey 
and briefly explain here (see also \cite{SESIET08,DESIsys24}  
for different variations).
First, the linear matter power spectrum~$P_\up{lin}(k)$ computed from 
the Boltzmann code \cite{CLASS} is decomposed 
into the \emph{wiggle} component~$P_\up{w}(k)$
and the \emph{no-wiggle} component~$P_\up{nw}(k)$:
\beeq
    P_\up{lin}(k) = P_\up{nw}(k) +  P_\up{w}(k)\,,
\eneq
where the broad-band shape of the no-wiggle power spectrum is
generated by using the fitting formula \cite{EIHU98}. 
As outlined in \cite*{ROBEET17},
we define the templates power spectrum~$P^t(k,\mu_k)$ for our BAO analysis as
\beeq
\label{templ}
P^{t}(k,\mu_k) =\left(\frac{1 + 
\beta \cdot \mu_k^2}{1+\mu_k^2 k^2 \Sigma_s^2 /2}\right)^2
\left[ P_\up{nw}(k) + e^{-k^2\sigma_v^2} \cdot P_\up{w}(k) \right]\,,
\eneq
where the prefactor in the round bracket accounts for the redshift-space
distortion with $\beta:=f/b$ and the Finger-of-god suppression with~$\Sigma_s$.
In addition, the wiggle power spectrum~$P_\up{w}(k)$ is further damped 
with~$\sigma_v$: 
\beeq
\sigma_v^2 = \frac{1}{2}\left[\left(1-\mu_k^2\right)\Sigma_\perp^2 + \mu_k^2 
\Sigma_{||}^2\right]~,
\eneq
to account for non-linear structure growth \cite{SESIET08,DESIsys24}
in redshift-space. As in \cite{ROBEET17}, we adopt the following
parameters
\beeq
 \Sigma_s = 4 h^{-1}\text{Mpc}~,\Dquad \beta = 0.76~, 
\Dquad  \Sigma_\perp = 6 h^{-1}\text{Mpc}~, \Dquad
\Sigma_{||} = 10 h^{-1}\text{Mpc}~.
\eneq

Given the template power spectrum~$P^t(k,\mu_k)$ in Eq.~\eqref{templ},
we first compute the template correlation function~$\xi^t(s,\mu)$ by
Fourier transformation. The template correlation function is then decomposed
in terms of two multipole correlation functions as
\beeq
    {\xi}^t_0(s,\alpha_{\perp},\alpha_{||}) = \int_0^{1} d\mu\;  \xi^t\left(s', 
\mu'\right)\,,\Tquad
    {\xi}^t_2(s,\alpha_{\perp},\alpha_{||}) = \int_0^{1} d\mu\; 3(\mu')^2~
 \xi^t\left(s', \mu'\right)\,,
\eneq
where the separation~$s'$ and angle~$\mu'$ for the fiducial cosmology
of the template are related to the 
observed separation~$s$ and angle~$\mu$ given the dilation parameters~$\al_\perp$
and~$\al_\para$ as 
\bear
&&
s'(s, \mu, \alpha_{\perp}, \alpha_{||}) = s \sqrt{\mu^2 \alpha_{||}^2 +
 (1 - \mu^2) \alpha_{\perp}^2}~, \\
&&
\mu'(\mu, \alpha_{\perp}, \alpha_{||}) = 
\frac{\mu \alpha_{||}}{\sqrt{\mu^2 \alpha_{||}^2 + (1 - \mu^2)
 \alpha_{\perp}^2}}\, . 
\enar
This procedure ensures that the template correlation function is properly
dilated for the cosmological model with one
represented by~$(\alpha_\perp,\alpha_\para)$.
Finally, the redshift-space multipoles correlation functions 
can be modeled through the template functions as
\bear
\label{BAOmono}
    \xi_{0}^\up{mod}(s,\alpha_{||},\alpha_{\perp}) &=&
 B_0 \cdot \xi^t_{0}(s,\alpha_{||},\alpha_{\perp}) + A_0(s)~,\\
    \xi_{2}^\up{mod}(s,\alpha_{||},\alpha_{\perp}) &=&
\label{BAOquad}
  \frac{5}{2} \left[B_2 \cdot \xi^t_{2}(s,\alpha_{||},\alpha_{\perp}) 
- B_0 \cdot \xi^t_{0}(s,\alpha_{||},\alpha_{\perp})\right] + A_2(s)~,
\enar
where two constants $B_x$ with $x \in \{0,2\}$ 
set the size of the BAO feature and $A_x(s) = a_{x,1}/s^2 + a_{x,2}/s + a_{x,3}$ 
account for the broad-band shape from any non-BAO contributions
\cite{SESIET08,ROBEET17}.
The redshift-space hexadecapole is not included in the analysis,
 as its  signal-to-noise around the BAO position is low given the current
survey volume \cite{VAHOET18}.

By combining the transverse and parallel dilation parameters, 
we obtain the isotropic dilation parameter
\begin{equation}
    \alpha_\up{iso}:=\big(\alpha_\perp^2\,\alpha_\parallel\big)^{1/3}=
\frac{D_V(z)r_d^{\mathrm{fid}}}{D_V^{\mathrm{fid}}(z)r_d}~,
\end{equation}
which provides constraints on the dimensionless ratio of $D_V(z)/r_d$~,
where we defined
\beeq
    D_V(z):=\Big[(1+z)^2 D_A^2(z)\,\frac{cz}{H(z)}\Big]^{1/3}~.
\eneq
The isotropic dilation parameter $\alpha_\up{iso}$ 
represents the distortion of the BAO
peak position due to the Alcock-Paczyński effect \cite{ALPA79}
as a geometric mean.

\subsection{Covariance Matrix for Redshift-Space Multipole
Correlation Functions}
\label{sec:cov}

The underlying fluctuations~$\delta(\kvec)$ are randomly distributed
around zero mean, following the Gaussian probability with the variance~$P(k)$.
Hence, the estimate $\hat P(\kvec)\simeq\delta(\kvec)\delta^*(\kvec)$
of the power spectrum also fluctuates around the mean value~$P(k)$ with
its variance. The covariance matrix for~$\hat P(\kvec)$ can be obtained
in the standard way \cite{FEKAPE94} for the redshift-space power spectrum,
and the covariance matrix per continuous
mode is then (see, e.g., \cite{GRSAET16,PESA16})
\beeq
\label{covcov}
\up{Cov}\bigg[\hat P(\kvec), ~\hat P(\kvec')\bigg]:
 = \langle \hat{P}(\mathbf{k})\hat{P}(\mathbf{k}') \rangle 
- \langle\hat{P}(\mathbf{k})\rangle  \langle\hat{P}(\mathbf{k}')\rangle\,
= \frac{2 (2\pi)^3 }{V_s}~ \delta^D({\mathbf{k - k'}}) 
\left[ P(\mathbf{k}) + \frac{1}{\bar{n}} \right]^2\,,
\eneq
where we subtracted the shot-noise contribution from the galaxy sample and
we have used the relation between the Kronecker delta~$\de^\up{K}_{k,k'}$ 
for a discrete mode
and the Dirac delta~$\de^D(\kvec-\kvec')$ functions for a continuous mode
in a given survey of 
volume~$V_s$:
\beeq
 \delta^\up{K}_{k,k'} = {(2\pi)^3\over V_s}~ \delta^D(\mathbf{k-k'})~.
\eneq
The covariance matrix of the redshift-space multipole power spectra~$P_l(k)$
defined in the same way for~$\hat P_l(k)$ as the covariance matrix
for~$\hat P(\kvec)$ in Eq.~\eqref{covcov}
can be directly computed by using the multipole decomposition relation in 
Eq.~\eqref{RSDDECOM}
\beeq
P_l(k)= \frac{2l + 1}{2} \int d \mu_k \ P(\kvec) \mathcal{L}_l(\mu_k)\,,
\eneq
as
\beeq
\up{Cov}\bigg[\hat P_l(k), ~\hat P_{l'}(k')\bigg]=
 \frac{2\pi^2}{k^2}~\delta^D({k-k'})
\frac{(2l + 1) (2l' + 1)}{V_s} \int d\mu_k \;\bigg[ P(\kvec) 
+ {1\over\bar n_g}\bigg]^2 \mathcal{L}_l(\mu_k) \mathcal{L}_{l'}(\mu_k) \; .
\eneq
Finally, we derive the covariance matrix of the redshift-space 
multipole correlation functions~$\xi_l(s)$ 
\beeq
\up{Cov}\bigg[\hat \xi_l(s), ~\hat \xi_{l'}(s')\bigg]
= c_{l,l'} \int d\ln k ~\frac{k^3}{2\pi^2} 
\bigg[ P(k) + {1\over\bar n_g} \bigg]^2 
\int d\mu_k ~ \mathcal{L}_l(\mu_k) \mathcal{L}_{l'}(\mu_k) j_l(ks)j_{l'}(ks')\; ,
\eneq
where we have defined the multipole coefficient
\beeq
c_{l,l'} := \frac{i^{l+l'}}{V_s}~(2l + 1) (2l' + 1) ~,
\eneq
and used the relation between the redshift-space multipoles
\beeq
\xi_l(s) = i^l \int d\ln k~{k^3\over2\pi^2}~ P_l(k) j_l(ks)\,.
\eneq

\section{Impact of the Monopole Corrections on the Galaxy Redshift-Space
Multipole-Correlation Functions and the BAO Peak Position}
\label{sec: Results}
Here we compute the corrections due to the monopole fluctuations on the
observed galaxy two-point correlation function and quantify the impact 
on BAO measurements.

\subsection{Multipole Correlation Functions and the Monopole Correction}
\label{sec:multipole}
Without a prior knowledge of the correct background mean number 
density~$\bar n_g(z)$, the angle average~$\bar n^\up{obs}_{g,\Omega}(z)$
is used to compute the observed galaxy fluctuation~$\de_g^\up{obs}(\Vang,z)$
in galaxy surveys, 
and the monopole fluctuation~$\de n(z)$ in~$\de_g^\up{obs}(\Vang,z)$
gives rise to the monopole correction $C_0(z_1,z_2)/4\pi$ in the
observed galaxy two-point correlation function in Eq.~\eqref{monocorr}.
Modeling the redshift-space galaxy number density 
fluctuation~$\de_g(\Vang,z)$ with the Kaiser formula in Eq.~\eqref{KAISER},
the monopole correction can be derived as
\beeq
\label{monoC0}
\frac{1}{4\pi} ~C_0 (z_1,z_2)
= D_1D_2 \int {d \ln{k}}\; \bigg[b_1 j_0(k r_1) - f_1 j_0 '' (k r_1)
 \bigg] \bigg[b_2 j_0(k r_2) - f_2 j_0 '' (k r_2) \bigg] \Delta_m^2(k)~,
\eneq
where the subscripts~$1,2$ for the growth factor~$D$, the bias factor~$b$, 
the logarithmic growth rate~$f$, and the comoving-distance~$r$,
represent redshifts~$z_1,z_2$ for its evaluation ($j_0$ is the spherical
Bessel function). Since the monopole fluctuation~$\de n(z)$ is just a function
of redshift, the monopole correction $C_0(z_1,z_2)/4\pi$ is a function of
two redshifts of the galaxy pair, independent of the pair separation and its
angle. However, since the galaxy correlation function~$\xi_g(s,\mu)$ is described
in terms of its separation~$s$ and angle~$\mu$, the monopole correction 
can also be described in terms of~$s$ and~$\mu$, and it displays
interesting dependence of~$s$ and~$\mu$, given that the mean redshift~$\bar z$
of the galaxy pair is fixed.

\begin{figure}[t]
\centering
\includegraphics[width=1\linewidth]{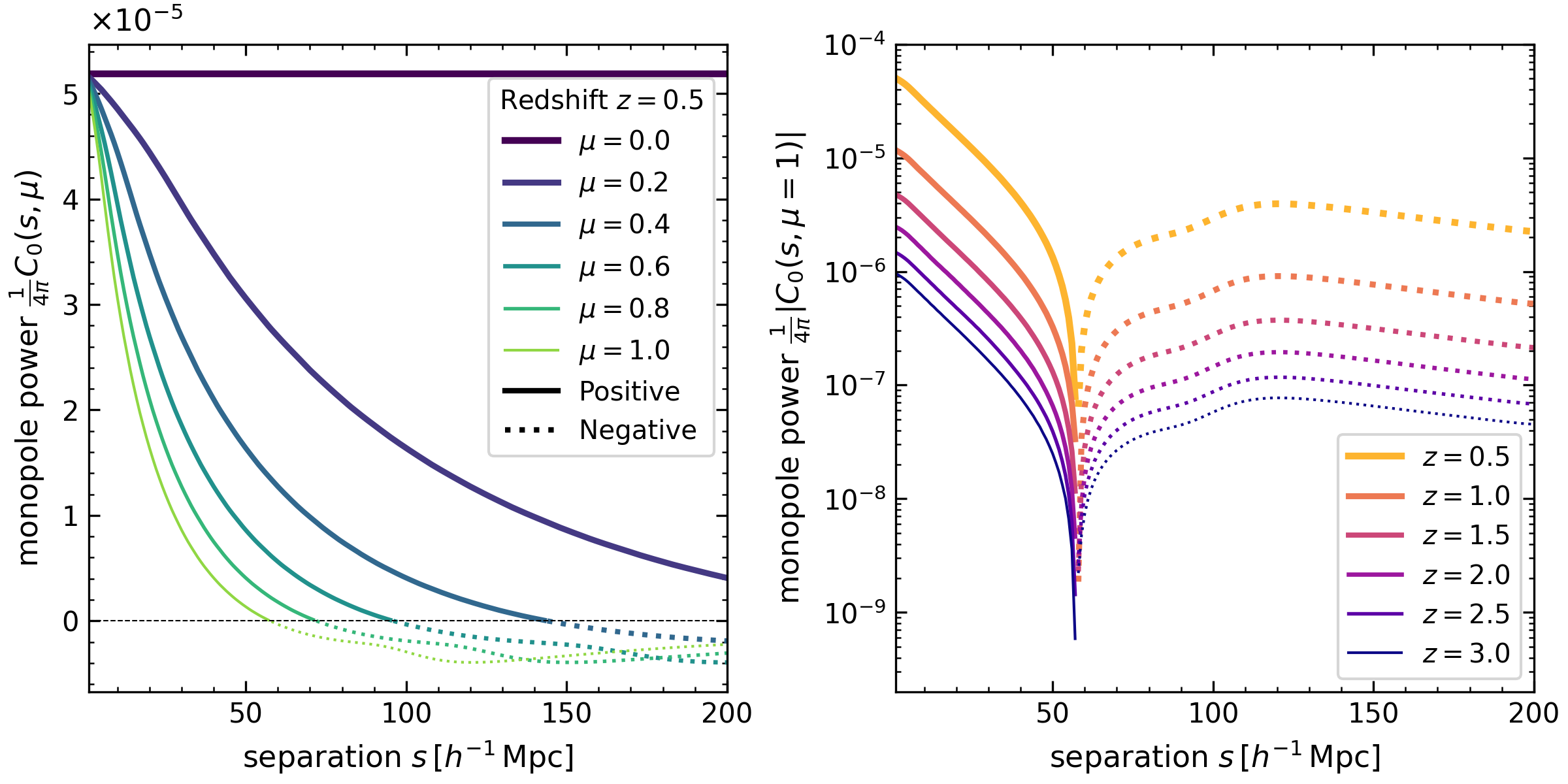}
\caption{Monopole correction $C_0/4\pi$ in the galaxy two-point correlation
function. With the dependence on two redshifts of the galaxy pair, the 
monopole correction is a function of pair separation~$s$ and angle~$\mu$. 
{\it Left panel}:
Various curves show the monopole correction with different angle~$\mu$.
The monopole correction is the same along the transverse 
direction ($\mu=0$), regardless of pair separation at the same redshift 
$z_1=z_2=0.5$. Along the line-of-sight direction ($\mu=1$), two galaxies are
located at different redshifts as a function of separation with the mean
redshift fixed at $\bar z=0.5$. {\it Right panel}: The monopole correction
for $\mu=1$ at different redshifts
is plotted in log scale as a function of separation.
With destructive interference of two spherical Bessel functions 
in Eq.~\eqref{monoC0}, the monopole correction is reduced and becomes negative,
compared to the monopole correction with constructive interference ($\mu=0$).}
\label{monopole-power}
\end{figure}

Figure~\ref{monopole-power} shows the monopole corrections~$C_0(s,\mu)/4\pi$ 
for galaxy pairs
in various configuration at various redshifts. We have assumed that the bias
factor is unity ($b\equiv1$). In the left panel, the thick horizontal line
shows the monopole correction for the galaxy pairs at the same redshift
$z_1=z_2=\bar z=0.5$. For this configuration ($\mu=0$), the monopole
correction is independent of the pair separation~$s$. However, for galaxy pairs
with $\mu\neq0$ (other curves), their redshifts~$z_1,z_2$ change 
as the pair separation~$s$ increases but with $\bar z=0.5$  fixed,
and hence the monopole correction changes. 
Of course,  in the limit the separation~$s\RA0$,
the monopole correction reduces to the value for~$\mu=0$.
As the difference between~$r_1$ and~$r_2$ increases, the interference
between two spherical Bessel functions becomes more destructive, and 
the monopole
correction decreases, even to negative values. This change is maximized 
along the line-of-sight direction ($\mu=1$). The right panel shows the
monopole correction in logarithm
at various redshifts along the line-of-sight direction.
At larger separations, where the monopole correction changes the sign
(dotted when negative), it reveals small wiggly structure due to the BAO 
feature in the matter power spectrum. 
The sign flip in~$C_0$ around the separation $s\simeq 55~\hmpc$ arises
independent of redshift. The dominant contribution to~$C_0/4\pi$ 
in Eq.~\eqref{monoC0} is in proportion to~$j_0(kr_1)j_0(kr_2)$ after
$j_0''(x)=-j_0(x)+2j_1(x)/x$ is substituted, and the product of two spherical
Bessel functions can be written in terms of a pair separation~$s$
and the mean distance~$\bar r$, instead of~$r_1$ and~$r_2$, showing that
the integral in Eq.~\eqref{monoC0} can be split as the redshift-dependent
amplitude (involving $D$, $b$, $f$, $\bar r$) multiplied
by the redshift-independent integration over~$k$, which 
involves $\Delta_m^2(k)$ and the pair separation~$s$. The redshift-independent
part determines the separation at which the monopole correction changes its
sign.

The  galaxy two-point correlation function~$\xi_g(s,\mu)$ 
is often analyzed in terms of their redshift-space multipole correlation 
function~$\xi_l(s)$ in Eq.~\eqref{ximul}. Since the observed galaxy
two-point correlation function~$\xi_g^\up{obs}(s,\mu)$ in Eq.~\eqref{monocorr}
contains the monopole correction, we decompose the monopole correction
in the same way by using the Legendre polynomials as
\beeq
\label{MONODEC}
\frac1{4\pi} C_0(s,\mu) =\sum_{l=\up{even}} \mathcal{L}_l(\mu) 
\mathcal{M}_l(s)\,,
\eneq
and the observed galaxy redshift-space multipole correlation function is then
\beeq
\xi^\up{obs}_{g,l} (s) = \xi_{l} (s) - \mathcal{M}_l(s)\,,\Dquad
\up{for}~~l=0,2,4,
\eneq
where the redshift-space multipole contribution~$\MMM_l(s)$ of the
monopole correction is defined as
\beeq
\label{MMMLLL}
  \mathcal{M}_l(s) := \frac{2l +1}{2}
\int_{-1}^{1} d\mu \, \mathcal{L}_l(\mu)~\frac1{4\pi} C_0(s,\mu)\,.
\eneq
Note that due to the symmetry $C_0(z_1,z_2)=C_0(z_2,z_1)$, the decomposition
in Eq.~\eqref{MONODEC} contains only the even multipoles. Though the 
decomposition of the monopole correction contains the multipoles 
beyond $l=0,2,4$, we only need $\MMM_l(s)$ with $l=0,2,4$.

\begin{figure}[t]
\centering
\includegraphics[width=1\linewidth]{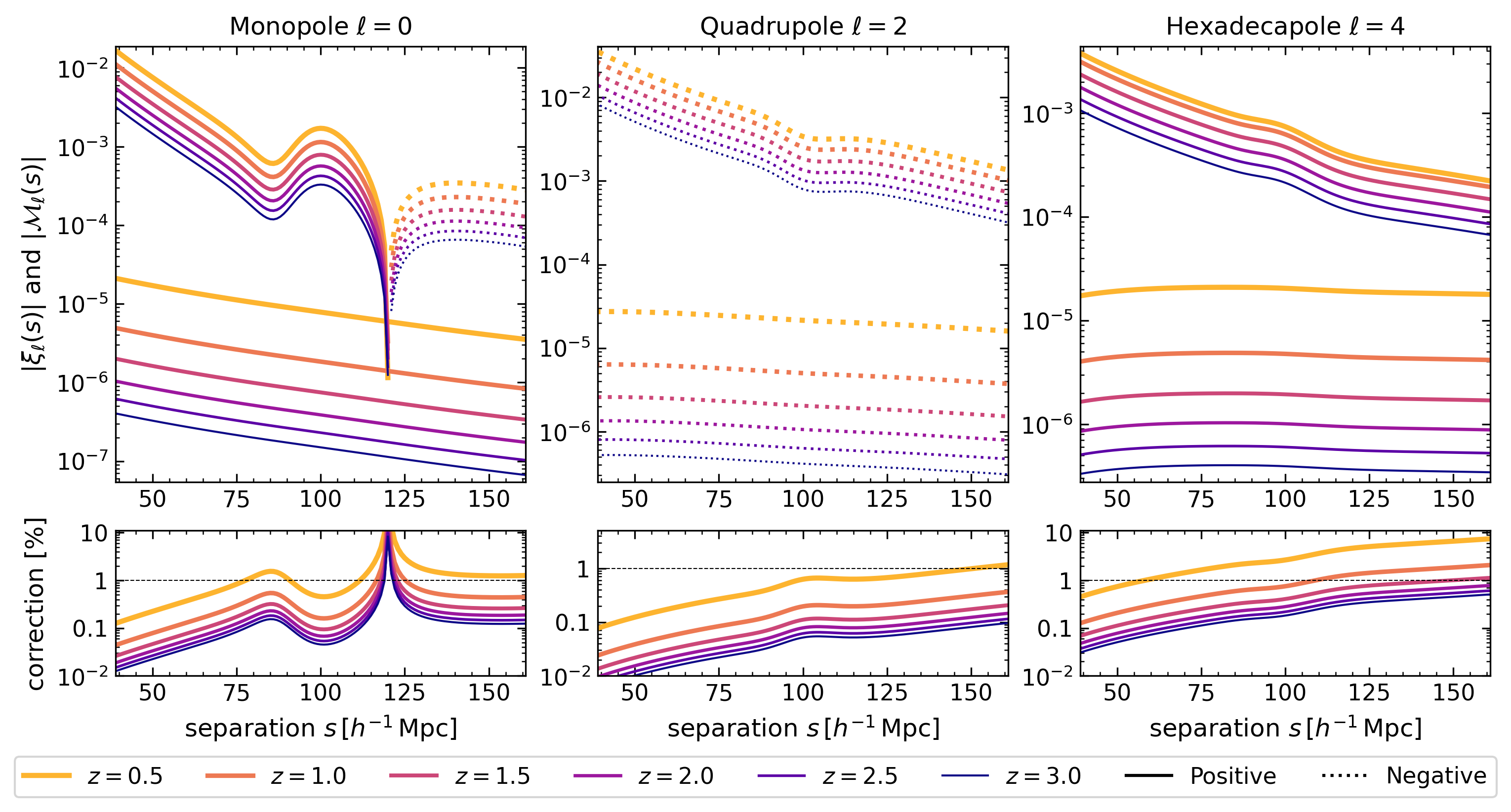}
\caption{Redshift-space multipole correlation functions~$\xi_l(s)$ 
and the multipole contributions~${\cal M}_l(s)$ of the monopole correction
$C_0(s,\mu)/4\pi$. As the monopole correction is a function of 
pair separation~$s$ and angle~$\mu$, it can be decomposed in terms of its
redshift-space multipole contributions~${\cal M}_l(s)$ in Eq.~\eqref{MONODEC}.
The upper panels show~$\xi_l(s)$ and~${\cal M}_l(s)$ for 
the monopole ({\it left}: $l=0$), the quadrupole ({\it middle}: $l=2$), 
the hexadecapole ({\it right}: $l=4$) at various redshifts (different color).
The bottom panels show the ratio ${\cal M}_l/\xi_l$ as a function of
separation~$s$. The monopole corrections
for the monopole and the quadrupole are at the sub-percent level around the 
BAO scale, while the correction is somewhat larger for the hexadecapole.
Though the ratio spikes around $\xi_0\simeq0$ in the left panel,
the multipole contributions~${\cal M}_l(s)$ of the monopole correction 
are always smooth around the BAO scale.}
\label{multipole}
\end{figure}

Figure~\ref{multipole} illustrates the redshift-space multipole correlation
functions~$\xi_l(s)$ and the multipole contributions~$\MMM_l(s)$ at various
redshifts. In the left panel, the monopole correlation function~$\xi_0(s)$
shows the BAO feature around $s\simeq100~\hmpc$, and it becomes negative
at larger separation. 
The multipole contribution~$\MMM_0(s)$ with $l=0$ of the monopole correction,
also shown in the upper panel, is the angle average of the monopole 
correction~$C_0(s,\mu)/4\pi$ in Figure~\ref{monopole-power} at a given 
separation~$s$. Once averaged, $\MMM_0(s)$ is a very smooth function of 
separation, and its ratio to~$\xi_0(s)$ is plotted in the bottom panel.
Overall, the multipole contribution~$\MMM_0(s)$ with $l=0$
is less than 1\% of~$\xi_0(s)$, 
while the ratio explodes around the scale ($\xi_0=0$), at which the
two-point correlation function becomes negative.

The middle and the right panels plot the multipole contributions~$\MMM_l(s)$
for quadrupole ($l=2$)
and hexadecapole ($l=4$), respectively. Compared to~$\xi_0(s)$, 
the quadrupole correlation function~$\xi_2(s)$ is more smooth,
and its BAO feature is much less pronounced. The 
 ratio of~$\MMM_2(s)$ to~$\xi_2(s)$ is similar to the
ratio in the monopole $(l=0)$, and the multipole contribution~$\MMM_2(s)$ is also
very smooth around the BAO scale. In comparison,
the multipole contribution~$\MMM_4(s)$
for the hexadecapole is much larger than the cases for the monopole and the
quadrupole. Overall, the ratios of~$\MMM_l(s)$ to~$\xi_l(s)$ with $l=0,2,4$
are larger at lower redshift, and they decrease fast at higher redshift.
While the ratio is independent of the growth factor~$D(z)$, the monopole
correction $C_0(z_1,z_2)/4\pi$ receives the dominant contributions from
the matter power spectrum~$\Delta^2_m(k)$ around $k\sim1/r_z$, and the
matter power spectrum decreases on larger scales, causing the
ratio $\MMM_l(s)/\xi_l(s)$ to decrease with redshift.

\begin{figure}[t]
\centering
\includegraphics[width=0.8\linewidth]{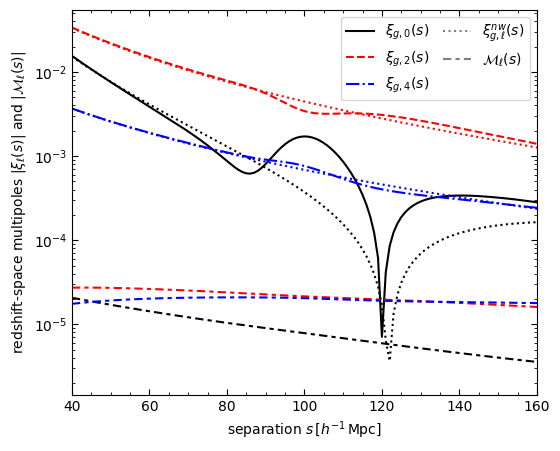}
\caption{Redshift-space multipole correlation functions~$\xi_l(s)$,
their broad-band no-wiggle correlation functions~$\xi_l^\up{nw}(s)$,
and the redshift-space multipole contributions~$\MMM_l(s)$ of the monopole
correction at $z=0.5$. Each redshift multipole is represented by
various color ($l=0$: black, $l=2$: red, $l=4$: blue). Dotted and short
dashed curves show~$\xi_l^\up{nw}(s)$ and~$\MMM_l(s)$.}
\label{shape}
\end{figure}

\subsection{BAO Peak Position Shift in the Monopole and Quadrupole
Correlation Functions}

The BAO peak position predicted by the linear theory differs from that
in the real observation. The nonlinear evolution of the matter and galaxy
bias and the redshift-space distortion effects
\cite{SESIET08,SMSCSH08,SEEKET10,YOMI10,SHZA12} are known to
broaden the BAO bump and shift the peak position below the sub-percent level.
However, these known systematic effects and 
unknown nonlinear power around the BAO
peak position can be modeled and marginalized over before identifying
the BAO peak position in the observed galaxy two-point correlation function.

Here we follow the BAO fitting procedure outlined in \cite{ROBEET17}
using the BAO model correlation functions~$\xi_0^\up{mod}(s)$
and~$\xi_2^\up{mod}(s)$ in Sec.~\ref{sec:BAO}.
The fitting procedure proceeds in two distinct steps. First, it focuses 
 on the monopole component~$\xi_0^{\text{mod}}(s)$ in Eq.~\eqref{BAOmono}
over the range \( s \in [40,160] \, h^{-1} \text{Mpc} \). We set
the broadband power $A_0(s)\equiv0$ in Eq.~\eqref{BAOmono} and 
allow only the amplitude parameter \( B_0 \) to vary
 to obtain an initial estimate of the amplitude~\( B_0 \), denoted 
as \( B_0^{\text{fit}} \), which will serve as a prior for the full 
parameter space exploration. Second, we extend the analysis to 
the complete set of parameters, considering ten free parameters: 
two dilation parameters~\( \alpha_{\perp} \) and~\( \alpha_{\parallel} \), 
two amplitude parameters~$B_0,B_2$, 
and six broadband correction parameters ~$a_{x,i}$ 
with $x \in \{0,2\}$ and $i \in \{1,2,3\}$ in Eq.~\eqref{BAOmono} and
Eq.~\eqref{BAOquad}.
To facilitate the parameter exploration, we impose a Gaussian prior 
on~$B_0$ and~$B_2$ centered in $B_0^{\text{fit}}$ with a 
width $\sigma_B = 0.4$ \cite{ROBEET17}. Note that $B_0=B_2$ is
expected in the fiducial model.

The observed galaxy redshift-space multipole correlation functions
are generated without any sample variance from the theoretical correlation
functions by assuming the fiducial cosmological parameters
\cite{PLANCKcos18,PLANCKover18},
and it includes the redshift-space multipole contributions~$\MMM_l(s)$
of the monopole correction. The data points are then sampled 
over the same range \( s \in [40,160] \, h^{-1} \text{Mpc} \)
in units of $1~\hmpc$, and 
we fit~$\xi_0^\up{mod}(s)$ and~$\xi_2^\up{mod}(s)$ to the observed 
galaxy redshift-space multipole correlation functions.
Furthermore, to compute the
covariance matrix of the measurements, we consider a galaxy sample
in a full-sky survey with the mean number density
$\bar n_g=5\times10^{-4}~(h/\up{Mpc})^3$,
the mean redshift~$\bar z=0.5$, and the redshift
depth $\Delta z=0.1$. The covariance matrix outlined in Sec.~\ref{sec:cov}
is computed by using the publicly available code \textsc{coffe} \cite{TAJEET18}.
For the parameter exploration, a Markov Chain Monte Carlo (MCMC) 
method using an affine-invariant ensemble sampler Python Package: 
\textsc{emcee} \cite{FOHOET13} was used.
We have performed ten MCMCs, setting 96 walkers with 12,000 steps, 
burn-in 90\%, with 10 different seeds to obtain the mean values of the
dilution parameters.

In conclusion, we find that the monopole fluctuations in the galaxy survey 
shift the BAO peak position by
\beeq
\alpha_\perp=1.0001~,\Tquad \alpha_\para=1.0002~,
\eneq
or in terms of the isotropic dilation parameter
\beeq
\al_\up{iso}=1.0002~.
\eneq
The best-fit values of the dilation parameters are the mean values
over ten MCMC realizations, and the errors on the mean are $\sim10^{-5}$.
The results indicate that the monopole fluctuations are almost
completely removed by the BAO fitting procedure and cause
{\it tiny} shift in the idealized case without any sample variance.
This level of shift in the BAO peak position is well 
below the level of measurement precision. The smoothness
of the monopole corrections in the redshift-space multipole correlation
functions are already manifest in Figure~\ref{multipole}. We show
in Figure~\ref{shape} the best-fit broadband power $\xi_{g,l}^\up{nw}(s)$
(dotted) and the redshift-space multipole contributions~$\MMM_l(s)$
(dashed) for the monopole ($l=0$; black), the quadrupole ($l=2$; red),
and the hexadecapole ($l=4$; blue). The redshift-space multipole
contribution~$\MMM_l(s)$ of the monopole correction~$C_0/4\pi$ is
much smaller than the broad-band power $\xi_{g,l}^\up{nw}(s)$, and
they are very smooth around the BAO peak to be absorbed into the 
power-law polynomials~$A_x(s)$ in Eqs.~\eqref{BAOmono} and~\eqref{BAOquad}.

\section{Conclusion and Discussion}
\label{sec: Conclusion}

The peak position of the baryonic acoustic oscillation (BAO) at each
redshift is a standard ruler that can be measured in galaxy surveys,
and it was shown \cite{SMSCSH08,SEEKET10,YOMI10,SHZA12}
that the BAO peak position in galaxy clustering
is robust against various systematic errors such as the redshift-space
distortion, galaxy bias,  the nonlinear evolution of matter, and the
gravitational lensing.
Given that the recent dark energy surveys such as the DESI plan
to achieve the sub-percent level of precision in the BAO analysis,
it is crucial to control any systematic errors below the targeted precision.
Here we have investigated the impact of a {\it new unknown} systematic error 
in the BAO analysis due to the monopole fluctuations. The monopole 
fluctuation~$\den(z)$ defined in Eq.~\eqref{angmon}
is the fluctuation in the observed galaxy
mean number density~$\ngavg(z)$ around the background mean number 
density~$\bar n_g(z)$ (see \cite{EIANET01,COEIET08,WHBLET11,DESIbao24}
for their estimates in galaxy surveys).
The galaxy number density fluctuation~$\dgobs(z,\Nang)$ 
is derived by subtracting the observed mean number density~$\ngavg(z)$
from the observed galaxy number density~$\ngobs(z,\Nang)$. Since the background
galaxy mean number density~$\bar n_g(z)$ is unknown, the observed mean
number density~$\ngavg(z)$ is obtained by averaging~$\ngobs(z,\Nang)$ 
over the angle,
which then includes the monopole fluctuation~$\den(z)$. It was shown
\cite{YOEI25} for the first time that the galaxy clustering
analysis such as the two-point and three-point correlation functions
is, therefore, affected  by the monopole fluctuation~$\den(z)$.
Arising from the galaxy number density fluctuation~$\de_g(z,\Nang)$ itself,
the monopole fluctuation is correlated with~$\de_g(z,\Nang)$.
Assuming a full sky galaxy survey, we have computed the redshift-space
multipole contributions~$\MMM_l(s)$ of the monopole correction~$C_0/4\pi$
to the redshift-space multipole correlation functions~$\xi_l(s)$
and estimated the shift~$\Delta \al$ in the BAO peak position due to the
un-modeled systematic error from the monopole correction by following
the standard procedure to marginalize over the broad-band power around 
the BAO peak and identify the shift parameter~$\Delta\al$.

While the monopole correction to the two-point correlation function
is only a function of two redshifts of the galaxy pair, it can be
expressed in terms of the pair separation and angle. 
The monopole correction shown in Figure~\ref{monopole-power}
is highly anisotropic,
and its amplitude is $5-10\%$ of the correlation function at the BAO
scale \cite{YOEI25}. However, once the monopole correction at a given
pair separation~$s$ is averaged over the angle, the multipole
contribution~$\MMM_l(s)$ in Eq.~\eqref{MMMLLL}
to the redshift-space multipole correlation
function~$\xi_l(s)$ turn out to be smooth over the BAO scale,
as illustrated in Figures~\ref{multipole} and~\ref{shape}.
While the relative amplitude shown in the
bottom panel is at the percent level, non-negligible given the level
of measurement precision, its smooth shape can be easily absorbed
into the broadband power in fitting the BAO template to the measurements
to isolate the shift in the BAO peak position.
To quantify the systematic errors in the BAO measurements,
we have adopted the standard BAO analysis \cite{ROBEET17}
(see also \cite{SESIET08,DESIsys24})
to estimate the shift in the BAO peak position due to the extra 
power~$\MMM_l(s)$ from the monopole fluctuations by fitting the BAO 
template to the
redshift-space monopole~$\xi_0(s)$ and quadrupole~$\xi_2(s)$ correlation
functions in the presence of the monopole fluctuations.
As already apparent in Figures~\ref{multipole} and~\ref{shape},
the redshift-space
multipole contributions~$\MMM_l(s)$ obtained by averaging the monopole
fluctuations over the angle are so smooth around the BAO peak in the
redshift-space multipole correlation functions that these contributions
are efficiently absorbed into the broad-band power that is designed
to account for any smooth power from the nonlinear evolution of matter
and galaxy bias  \cite{SMSCSH08,SEEKET10,SHZA12}.
 We find that the shift in the BAO peak position
due to the monopole fluctuations is $\Delta\al\approx10^{-4}$ 
in terms of the isotropic shift parameter and hence negligible for the current
dark energy surveys. Our results demonstrate that the current BAO analysis
is robust against the systematic error from the monopole fluctuations.

Throughout the manuscript, we have ignored the nonlinear contribution 
of the matter fluctuations. While the BAO scale is largely in the linear regime,
the correlation functions are affected by the nonlinear evolution of matter.
 For example, the monopole correlation function is reduced by 10\% at $z=0$ 
(the peak is less sharp) and 2\% at $z=0.5$ around the BAO scale 
by using the nonlinear matter power spectrum \cite{MAFR00,SMPEET03,HEFRET21}.
The nonlinearity effects, however, affect both the redshift-space multipole
correlation functions~$\xi_l(s)$ and the redshift-space multipole
contributions~$\MMM_l(s)$ in almost the same way, since the monopole
fluctuation arises directly from the galaxy number density fluctuation.
Hence,  the level of the monopole correction in the redshift-space multipole 
correlation functions remains unchanged. 
The same arguments can be made for the galaxy bias factor. We have assumed
that the galaxy distribution is the same as the matter fluctuation ($b=1$)
around the BAO scale, and we have considered only the linear bias factor,
ignoring the nonlinear bias factor such as~$b_2$ \cite{SZALA88,FRGA93,FRY96}.
In the same spirit, any change in the galaxy bias factor affects
both~$\xi_l(s)$ and~$\MMM_l(s)$ in almost the same way, and the level
of the monopole correction remains the same in the redshift-space
multipole correlation functions. Therefore, we suspect that our conclusion
on the impact of the monopole fluctuations would be insensitive to
the nonlinearity of the matter density fluctuations and the galaxy bias factor.
In our theoretical modeling of the observed galaxy two-point correlation
function, we have adopted the distant observer approximation and ignored
the wide angle effects \cite{SZMALA98,MATSU00,SZAPU04,PASZ08,YOSE15}. 
While the wide angle effects at the BAO scale are
important at lower redshift $z\ll0.1$, it can be safely ignored at higher
redshift. 

The monopole fluctuations and the monopole corrections in the
present work are computed by assuming all-sky galaxy surveys ($\Omega=4\pi$). 
In practice,
galaxy surveys cover only a small fraction of the sky, for example, 
24\% for BOSS and 34\% for DESI. The total sky coverage~$\Omega$ 
of galaxy surveys is
indeed composed of several disjoint regions, which makes the effective
sky coverage even smaller in terms of estimating the observed galaxy mean
number density~$\bar n_{g,\Omega}^\up{obs}(z)$ in Eq.~\eqref{NBOBSOM}.
When the sky coverage is smaller, the other angular multipole moments~$a_{lm}$
with $l>0$ in the galaxy number density fluctuation~$\de_g(\Vang,z)$
contribute to the ``monopole fluctuation'' $\de n(z)$ in Eq.~\eqref{angmon}
as these low angular multipoles are smooth over the small patch of the sky
and indistinguishable from the mean background number density~$\bar n_g(z)$. 
Hence, the monopole fluctuation~$\de n(z)$, though its name in this case is a
misnomer, is greatly enhanced in terms of its amplitude and correlation,
compared to the same galaxy fluctuation~$\de_g(\Vang,z)$.
We defer this follow-up investigation for the future work \cite{PIMAET26}.

\section*{Acknowledgments}
We acknowledge useful discussions with Yan-Chuan Cai.
This work is support by the Swiss National Science Foundation.

\bibliography{paper.bbl}

\providecommand{\href}[2]{#2}\begingroup\raggedright\begin{thebibliography}{10}

\bibitem{EIHU98}
D.~J. {Eisenstein} and W.~{Hu}, \emph{{Baryonic Features in the Matter Transfer
  Function}}, \href{http://dx.doi.org/10.1086/305424}{\emph{\apj} {\bfseries
  496} (Mar., 1998) 605},
  [\href{https://arxiv.org/abs/arXiv:astro-ph/9709112}{{\ttfamily
  arXiv:astro-ph/9709112}}].

\bibitem{EISEWH07}
D.~J. {Eisenstein}, H.-J. {Seo} and M.~{White}, \emph{{On the Robustness of the
  Acoustic Scale in the Low-Redshift Clustering of Matter}}, {\emph{\apj}
  {\bfseries 664} (Aug., 2007) 660--674},
  [\href{https://arxiv.org/abs/arXiv:astro-ph/0604361}{{\ttfamily
  arXiv:astro-ph/0604361}}].

\bibitem{WEMOET13}
D.~H. {Weinberg}, M.~J. {Mortonson}, D.~J. {Eisenstein}, C.~{Hirata}, A.~G.
  {Riess} and E.~{Rozo}, \emph{{Observational probes of cosmic acceleration}},
  \href{http://dx.doi.org/10.1016/j.physrep.2013.05.001}{\emph{\physrep}
  {\bfseries 530} (Sept., 2013) 87--255},
  [\href{https://arxiv.org/abs/1201.2434}{{\ttfamily 1201.2434}}].

\bibitem{PEYU70}
P.~J.~E. {Peebles} and J.~T. {Yu}, \emph{{Primeval Adiabatic Perturbation in an
  Expanding Universe}}, {\emph{\apj} {\bfseries 162} (Dec., 1970) 815}.

\bibitem{SUZE70a}
R.~A. {Sunyaev} and Y.~B. {Zeldovich}, \emph{{The Spectrum of Primordial
  Radiation, its Distortions and their Significance}}, {\emph{Ap\&SS}
  {\bfseries 2} (Mar., 1970) 66}.

\bibitem{PLANCKcos18}
{Planck Collaboration}, N.~{Aghanim}, Y.~{Akrami}, M.~{Ashdown}, J.~{Aumont},
  C.~{Baccigalupi} et~al., \emph{{Planck 2018 results. VI. Cosmological
  parameters}},
  \href{http://dx.doi.org/10.1051/0004-6361/201833910}{\emph{\aap} {\bfseries
  641} (Sept., 2020) A6}, [\href{https://arxiv.org/abs/1807.06209}{{\ttfamily
  1807.06209}}].

\bibitem{PLANCKover18}
{Planck Collaboration}, N.~{Aghanim}, Y.~{Akrami}, F.~{Arroja}, M.~{Ashdown},
  J.~{Aumont} et~al., \emph{{Planck 2018 results. I. Overview and the
  cosmological legacy of Planck}}, {\emph{arXiv e-prints} (July, 2018) },
  [\href{https://arxiv.org/abs/1807.06205}{{\ttfamily 1807.06205}}].

\bibitem{YOADET00}
D.~G. {York}, J.~{Adelman}, J.~{Anderson}, John~E., S.~F. {Anderson},
  J.~{Annis}, N.~A. {Bahcall} et~al., \emph{{The Sloan Digital Sky Survey:
  Technical Summary}}, \href{http://dx.doi.org/10.1086/301513}{\emph{\aj}
  {\bfseries 120} (Sept., 2000) 1579--1587},
  [\href{https://arxiv.org/abs/arXiv:astro-ph/0006396}{{\ttfamily
  arXiv:astro-ph/0006396}}].

\bibitem{EIZEET05}
D.~J. {Eisenstein}, I.~{Zehavi}, D.~W. {Hogg}, R.~{Scoccimarro}, M.~R.
  {Blanton}, R.~C. {Nichol} et~al., \emph{{Detection of the Baryon Acoustic
  Peak in the Large-Scale Correlation Function of SDSS Luminous Red Galaxies}},
  {\emph{\apj} {\bfseries 633} (Nov., 2005) 560--574},
  [\href{https://arxiv.org/abs/arXiv:astro-ph/0501171}{{\ttfamily
  arXiv:astro-ph/0501171}}].

\bibitem{BEBLET11}
F.~{Beutler}, C.~{Blake}, M.~{Colless}, D.~H. {Jones}, L.~{Staveley-Smith},
  L.~{Campbell} et~al., \emph{{The 6dF Galaxy Survey: baryon acoustic
  oscillations and the local Hubble constant}}, {\emph{\mnras} {\bfseries 416}
  (Oct., 2011) 3017--3032}, [\href{https://arxiv.org/abs/1106.3366}{{\ttfamily
  1106.3366}}].

\bibitem{BLBRET11}
C.~{Blake}, S.~{Brough} et~al., \emph{{The WiggleZ Dark Energy Survey: the
  selection function and z = 0.6 galaxy power spectrum}}, {\emph{\mnras}
  {\bfseries 406} (Aug., 2010) 803--821},
  [\href{https://arxiv.org/abs/1003.5721}{{\ttfamily 1003.5721}}].

\bibitem{DESI13}
M.~{Levi}, C.~{Bebek}, T.~{Beers}, R.~{Blum}, R.~{Cahn}, D.~{Eisenstein}
  et~al., \emph{{The DESI Experiment, a whitepaper for Snowmass 2013}},
  {\emph{ArXiv e-prints} (Aug., 2013) },
  [\href{https://arxiv.org/abs/1308.0847}{{\ttfamily 1308.0847}}].

\bibitem{BLCOET07}
C.~{Blake}, A.~{Collister}, S.~{Bridle} and O.~{Lahav}, \emph{{Cosmological
  baryonic and matter densities from 600000 SDSS luminous red galaxies with
  photometric redshifts}}, {\emph{\mnras} {\bfseries 374} (Feb., 2007)
  1527--1548}, [\href{https://arxiv.org/abs/arXiv:astro-ph/0605303}{{\ttfamily
  arXiv:astro-ph/0605303}}].

\bibitem{PEREET10}
W.~J. {Percival}, B.~A. {Reid}, D.~J. {Eisenstein}, N.~A. {Bahcall},
  T.~{Budavari}, J.~A. {Frieman} et~al., \emph{{Baryon acoustic oscillations in
  the Sloan Digital Sky Survey Data Release 7 galaxy sample}}, {\emph{\mnras}
  {\bfseries 401} (Feb., 2010) 2148--2168},
  [\href{https://arxiv.org/abs/0907.1660}{{\ttfamily 0907.1660}}].

\bibitem{ANAUET12}
L.~{Anderson}, E.~{Aubourg}, S.~{Bailey}, D.~{Bizyaev}, M.~{Blanton}, A.~S.
  {Bolton} et~al., \emph{{The clustering of galaxies in the SDSS-III Baryon
  Oscillation Spectroscopic Survey: baryon acoustic oscillations in the Data
  Release 9 spectroscopic galaxy sample}}, {\emph{\mnras} {\bfseries 427}
  (Dec., 2012) 3435--3467}, [\href{https://arxiv.org/abs/1203.6594}{{\ttfamily
  1203.6594}}].

\bibitem{ATBAET18}
M.~{Ata}, F.~{Baumgarten}, J.~{Bautista}, F.~{Beutler}, D.~{Bizyaev}, M.~R.
  {Blanton} et~al., \emph{{The clustering of the SDSS-IV extended Baryon
  Oscillation Spectroscopic Survey DR14 quasar sample: first measurement of
  baryon acoustic oscillations between redshift 0.8 and 2.2}},
  \href{http://dx.doi.org/10.1093/mnras/stx2630}{\emph{\mnras} {\bfseries 473}
  (Feb., 2018) 4773--4794}, [\href{https://arxiv.org/abs/1705.06373}{{\ttfamily
  1705.06373}}].

\bibitem{BAPAET21}
J.~E. {Bautista}, R.~{Paviot}, M.~{Vargas Maga{\~n}a}, S.~{de la Torre},
  S.~{Fromenteau}, H.~{Gil-Mar{\'\i}n} et~al., \emph{{The completed SDSS-IV
  extended Baryon Oscillation Spectroscopic Survey: measurement of the BAO and
  growth rate of structure of the luminous red galaxy sample from the
  anisotropic correlation function between redshifts 0.6 and 1}},
  \href{http://dx.doi.org/10.1093/mnras/staa2800}{\emph{\mnras} {\bfseries 500}
  (Jan., 2021) 736--762}, [\href{https://arxiv.org/abs/2007.08993}{{\ttfamily
  2007.08993}}].

\bibitem{DESIbao24}
{DESI Collaboration}, A.~G. {Adame}, J.~{Aguilar}, S.~{Ahlen}, S.~{Alam}, D.~M.
  {Alexander} et~al., \emph{{DESI 2024 VI: Cosmological Constraints from the
  Measurements of Baryon Acoustic Oscillations}},
  \href{http://dx.doi.org/10.48550/arXiv.2404.03002}{\emph{arXiv e-prints}
  (Apr., 2024) }, [\href{https://arxiv.org/abs/2404.03002}{{\ttfamily
  2404.03002}}].

\bibitem{DESIcos24}
{DESI Collaboration}, A.~G. {Adame}, J.~{Aguilar}, S.~{Ahlen}, S.~{Alam}, D.~M.
  {Alexander} et~al., \emph{{DESI 2024 VI: Cosmological Constraints from the
  Measurements of Baryon Acoustic Oscillations}},
  \href{http://dx.doi.org/10.48550/arXiv.2404.03002}{\emph{arXiv e-prints}
  (Apr., 2024) }, [\href{https://arxiv.org/abs/2404.03002}{{\ttfamily
  2404.03002}}].

\bibitem{YOEI25}
J.~{Yoo} and D.~{Eisenstein}, \emph{{Monopole Fluctuations in Galaxy Surveys}},
  \href{http://dx.doi.org/10.3847/2041-8213/ada911}{\emph{\apjl} {\bfseries
  979} (Feb., 2025) L35}, [\href{https://arxiv.org/abs/2410.00951}{{\ttfamily
  2410.00951}}].

\bibitem{ROBEET17}
A.~J. {Ross}, F.~{Beutler}, C.-H. {Chuang}, M.~{Pellejero-Ibanez}, H.-J. {Seo},
  M.~{Vargas-Maga{\~n}a} et~al., \emph{{The clustering of galaxies in the
  completed SDSS-III Baryon Oscillation Spectroscopic Survey: observational
  systematics and baryon acoustic oscillations in the correlation function}},
  \href{http://dx.doi.org/10.1093/mnras/stw2372}{\emph{\mnras} {\bfseries 464}
  (Jan., 2017) 1168--1191}, [\href{https://arxiv.org/abs/1607.03145}{{\ttfamily
  1607.03145}}].

\bibitem{PEEBL80}
P.~J.~E. {Peebles}, \emph{{The large-scale structure of the universe}}.
\newblock Princeton University Press, Princeton, 1980.

\bibitem{PEACO99}
J.~A. {Peacock}, \emph{{Cosmological Physics}}.
\newblock Cambridge University Press, January 1999.

\bibitem{DODEL03}
S.~{Dodelson}, \emph{{Modern cosmology}}.
\newblock Academic Press., 2003.

\bibitem{KAISE87}
N.~{Kaiser}, \emph{{Clustering in real space and in redshift space}},
  {\emph{\mnras} {\bfseries 227} (July, 1987) 1--21}.

\bibitem{KAISE84}
N.~{Kaiser}, \emph{{On the spatial correlations of Abell clusters}},
  {\emph{\apjl} {\bfseries 284} (Sept., 1984) L9--L12}.

\bibitem{BBKS86}
J.~M. {Bardeen}, J.~R. {Bond}, N.~{Kaiser} and A.~S. {Szalay}, \emph{{The
  statistics of peaks of Gaussian random fields}},
  \href{http://dx.doi.org/10.1086/164143}{\emph{\apj} {\bfseries 304} (May,
  1986) 15--61}.

\bibitem{YOO09}
J.~{Yoo}, \emph{{Complete Treatment of Galaxy Two-Point Statistics:
  Gravitational Lensing Effects and Redshift-Space Distortions}},
  \href{http://dx.doi.org/10.1103/PhysRevD.79.023517}{\emph{\prd} {\bfseries
  79} (Jan., 2009) 023517},
  [\href{https://arxiv.org/abs/arXiv:0808.3138}{{\ttfamily arXiv:0808.3138}}].

\bibitem{YOFIZA09}
J.~{Yoo}, A.~L. {Fitzpatrick} and M.~{Zaldarriaga}, \emph{{New perspective on
  galaxy clustering as a cosmological probe: General relativistic effects}},
  \href{http://dx.doi.org/10.1103/PhysRevD.80.083514}{\emph{\prd} {\bfseries
  80} (Oct., 2009) 083514},
  [\href{https://arxiv.org/abs/arXiv:0907.0707}{{\ttfamily arXiv:0907.0707}}].

\bibitem{YOO14a}
J.~{Yoo}, \emph{{Relativistic effect in galaxy clustering}},
  \href{http://dx.doi.org/10.1088/0264-9381/31/23/234001}{\emph{\cqg}
  {\bfseries 31} (Sept., 2014) 234001},
  [\href{https://arxiv.org/abs/arXiv:1409.3223}{{\ttfamily arXiv:1409.3223}}].

\bibitem{HAMIL92}
A.~J.~S. {Hamilton}, \emph{{Measuring Omega and the real correlation function
  from the redshift correlation function}}, {\emph{\apjl} {\bfseries 385}
  (Jan., 1992) L5--L8}.

\bibitem{COFIWE94}
S.~{Cole}, K.~B. {Fisher} and D.~H. {Weinberg}, \emph{{Fourier Analysis of
  Redshift Space Distortions and the Determination of Omega}}, {\emph{\mnras}
  {\bfseries 267} (Apr., 1994) 785},
  [\href{https://arxiv.org/abs/arXiv:astro-ph/9308003}{{\ttfamily
  arXiv:astro-ph/9308003}}].

\bibitem{HAMAET03}
E.~{Hawkins}, S.~{Maddox}, S.~{Cole}, O.~{Lahav}, D.~S. {Madgwick},
  P.~{Norberg} et~al., \emph{{The 2dF Galaxy Redshift Survey: correlation
  functions, peculiar velocities and the matter density of the Universe}},
  \href{http://dx.doi.org/10.1046/j.1365-2966.2003.07063.x}{\emph{\mnras}
  {\bfseries 346} (Nov., 2003) 78--96},
  [\href{https://arxiv.org/abs/astro-ph/0212375}{{\ttfamily
  astro-ph/0212375}}].

\bibitem{RESAET12}
B.~A. {Reid}, L.~{Samushia}, M.~{White}, W.~J. {Percival}, M.~{Manera},
  N.~{Padmanabhan} et~al., \emph{{The clustering of galaxies in the SDSS-III
  Baryon Oscillation Spectroscopic Survey: measurements of the growth of
  structure and expansion rate at z = 0.57 from anisotropic clustering}},
  \href{http://dx.doi.org/10.1111/j.1365-2966.2012.21779.x}{\emph{\mnras}
  {\bfseries 426} (Nov., 2012) 2719--2737},
  [\href{https://arxiv.org/abs/1203.6641}{{\ttfamily 1203.6641}}].

\bibitem{BESEET17}
F.~{Beutler}, H.-J. {Seo}, S.~{Saito}, C.-H. {Chuang}, A.~J. {Cuesta}, D.~J.
  {Eisenstein} et~al., \emph{{The clustering of galaxies in the completed
  SDSS-III Baryon Oscillation Spectroscopic Survey: anisotropic galaxy
  clustering in Fourier space}},
  \href{http://dx.doi.org/10.1093/mnras/stw3298}{\emph{\mnras} {\bfseries 466}
  (Apr., 2017) 2242--2260}, [\href{https://arxiv.org/abs/1607.03150}{{\ttfamily
  1607.03150}}].

\bibitem{EIANET01}
D.~J. {Eisenstein}, J.~{Annis}, J.~E. {Gunn}, A.~S. {Szalay}, A.~J. {Connolly},
  R.~C. {Nichol} et~al., \emph{{Spectroscopic Target Selection for the Sloan
  Digital Sky Survey: The Luminous Red Galaxy Sample}}, {\emph{\aj} {\bfseries
  122} (Nov., 2001) 2267--2280},
  [\href{https://arxiv.org/abs/arXiv:astro-ph/0108153}{{\ttfamily
  arXiv:astro-ph/0108153}}].

\bibitem{COEIET08}
R.~J. {Cool}, D.~J. {Eisenstein}, X.~{Fan}, M.~{Fukugita}, L.~{Jiang},
  C.~{Maraston} et~al., \emph{{Luminosity Function Constraints on the Evolution
  of Massive Red Galaxies since z \~{} 0.9}}, {\emph{\apj} {\bfseries 682}
  (Aug., 2008) 919--936}, [\href{https://arxiv.org/abs/0804.4516}{{\ttfamily
  0804.4516}}].

\bibitem{WHBLET11}
M.~{White}, M.~{Blanton}, A.~{Bolton}, D.~{Schlegel}, J.~{Tinker}, A.~{Berlind}
  et~al., \emph{{The Clustering of Massive Galaxies at z \~{} 0.5 from the
  First Semester of BOSS Data}}, {\emph{\apj} {\bfseries 728} (Feb., 2011)
  126}, [\href{https://arxiv.org/abs/1010.4915}{{\ttfamily 1010.4915}}].

\bibitem{MIYOET20}
E.~{Mitsou}, J.~{Yoo}, R.~{Durrer}, F.~{Scaccabarozzi} and V.~{Tansella},
  \emph{{General and consistent statistics for cosmological observations}},
  \href{http://dx.doi.org/10.1103/PhysRevResearch.2.033004}{\emph{Physical
  Review Research} {\bfseries 2} (July, 2020) 033004},
  [\href{https://arxiv.org/abs/1905.01293}{{\ttfamily 1905.01293}}].

\bibitem{EIHUET98}
D.~J. {Eisenstein}, W.~{Hu}, J.~{Silk} and A.~S. {Szalay}, \emph{{Can Baryonic
  Features Produce the Observed 100 h$^{-1}$ Mpc Clustering?}},
  \href{http://dx.doi.org/10.1086/311171}{\emph{\apjl} {\bfseries 494} (Feb.,
  1998) L1--L4}, [\href{https://arxiv.org/abs/astro-ph/9710303}{{\ttfamily
  astro-ph/9710303}}].

\bibitem{EIHU99}
D.~J. {Eisenstein} and W.~{Hu}, \emph{{Power Spectra for Cold Dark Matter and
  Its Variants}}, {\emph{\apj} {\bfseries 511} (Jan., 1999) 5--15},
  [\href{https://arxiv.org/abs/astro-ph/9710252}{{\ttfamily
  astro-ph/9710252}}].

\bibitem{EIWH04}
D.~{Eisenstein} and M.~{White}, \emph{{Theoretical uncertainty in baryon
  oscillations}}, {\emph{\prd} {\bfseries 70} (Nov., 2004) 103523},
  [\href{https://arxiv.org/abs/arXiv:astro-ph/0407539}{{\ttfamily
  arXiv:astro-ph/0407539}}].

\bibitem{ALPA79}
C.~{Alcock} and B.~{Paczy\'nski}, \emph{{An evolution free test for non-zero
  cosmological constant}}, {\emph{\nat} {\bfseries 281} (Oct., 1979) 358}.

\bibitem{SESIET08}
H.-J. {Seo}, E.~R. {Siegel}, D.~J. {Eisenstein} and M.~{White},
  \emph{{Nonlinear Structure Formation and the Acoustic Scale}},
  \href{http://dx.doi.org/10.1086/589921}{\emph{\apj} {\bfseries 686} (Oct.,
  2008) 13--24}, [\href{https://arxiv.org/abs/0805.0117}{{\ttfamily
  0805.0117}}].

\bibitem{DESIsys24}
S.-F. {Chen}, C.~{Howlett}, M.~{White}, P.~{McDonald}, A.~J. {Ross}, H.-J.
  {Seo} et~al., \emph{{Baryon Acoustic Oscillation Theory and Modelling
  Systematics for the DESI 2024 results}},
  \href{http://dx.doi.org/10.48550/arXiv.2402.14070}{\emph{arXiv e-prints}
  (Feb., 2024) arXiv:2402.14070},
  [\href{https://arxiv.org/abs/2402.14070}{{\ttfamily 2402.14070}}].

\bibitem{CLASS}
D.~{Blas}, J.~{Lesgourgues} and T.~{Tram}, \emph{{The Cosmic Linear Anisotropy
  Solving System (CLASS). Part II: Approximation schemes}},
  \href{http://dx.doi.org/10.1088/1475-7516/2011/07/034}{\emph{\jcap}
  {\bfseries 7} (July, 2011) 034},
  [\href{https://arxiv.org/abs/1104.2933}{{\ttfamily 1104.2933}}].

\bibitem{VAHOET18}
M.~{Vargas-Maga{\~n}a}, S.~{Ho}, A.~J. {Cuesta}, R.~{O'Connell}, A.~J. {Ross},
  D.~J. {Eisenstein} et~al., \emph{{The clustering of galaxies in the completed
  SDSS-III Baryon Oscillation Spectroscopic Survey: theoretical systematics and
  Baryon Acoustic Oscillations in the galaxy correlation function}},
  \href{http://dx.doi.org/10.1093/mnras/sty571}{\emph{\mnras} {\bfseries 477}
  (June, 2018) 1153--1188}, [\href{https://arxiv.org/abs/1610.03506}{{\ttfamily
  1610.03506}}].

\bibitem{FEKAPE94}
H.~A. {Feldman}, N.~{Kaiser} and J.~A. {Peacock}, \emph{{Power-spectrum
  analysis of three-dimensional redshift surveys}}, {\emph{\apj} {\bfseries
  426} (May, 1994) 23--37},
  [\href{https://arxiv.org/abs/arXiv:astro-ph/9304022}{{\ttfamily
  arXiv:astro-ph/9304022}}].

\bibitem{GRSAET16}
J.~N. {Grieb}, A.~G. {S{\'a}nchez}, S.~{Salazar-Albornoz} and C.~{Dalla
  Vecchia}, \emph{{Gaussian covariance matrices for anisotropic galaxy
  clustering measurements}},
  \href{http://dx.doi.org/10.1093/mnras/stw065}{\emph{\mnras} {\bfseries 457}
  (Apr., 2016) 1577--1592}, [\href{https://arxiv.org/abs/1509.04293}{{\ttfamily
  1509.04293}}].

\bibitem{PESA16}
D.~W. {Pearson} and L.~{Samushia}, \emph{{Estimating the power spectrum
  covariance matrix with fewer mock samples}},
  \href{http://dx.doi.org/10.1093/mnras/stw062}{\emph{\mnras} {\bfseries 457}
  (Mar., 2016) 993--999}, [\href{https://arxiv.org/abs/1509.00064}{{\ttfamily
  1509.00064}}].

\bibitem{SMSCSH08}
R.~E. {Smith}, R.~{Scoccimarro} and R.~K. {Sheth}, \emph{{Motion of the
  acoustic peak in the correlation function}},
  \href{http://dx.doi.org/10.1103/PhysRevD.77.043525}{\emph{\prd} {\bfseries
  77} (Feb., 2008) 043525},
  [\href{https://arxiv.org/abs/arXiv:astro-ph/0703620}{{\ttfamily
  arXiv:astro-ph/0703620}}].

\bibitem{SEEKET10}
H.-J. {Seo}, J.~{Eckel}, D.~J. {Eisenstein}, K.~{Mehta}, M.~{Metchnik},
  N.~{Padmanabhan} et~al., \emph{{High-precision Predictions for the Acoustic
  Scale in the Nonlinear Regime}}, {\emph{\apj} {\bfseries 720} (Sept., 2010)
  1650--1667}, [\href{https://arxiv.org/abs/0910.5005}{{\ttfamily 0910.5005}}].

\bibitem{YOMI10}
J.~{Yoo} and J.~{Miralda-Escud{\'e}}, \emph{{Gravitational lensing effects on
  the baryonic acoustic oscillation signature in the redshift-space correlation
  function}}, \href{http://dx.doi.org/10.1103/PhysRevD.82.043527}{\emph{\prd}
  {\bfseries 82} (Aug., 2010) 043527},
  [\href{https://arxiv.org/abs/0901.0708}{{\ttfamily 0901.0708}}].

\bibitem{SHZA12}
B.~D. {Sherwin} and M.~{Zaldarriaga}, \emph{{Shift of the baryon acoustic
  oscillation scale: A simple physical picture}},
  \href{http://dx.doi.org/10.1103/PhysRevD.85.103523}{\emph{\prd} {\bfseries
  85} (May, 2012) 103523}, [\href{https://arxiv.org/abs/1202.3998}{{\ttfamily
  1202.3998}}].

\bibitem{TAJEET18}
V.~{Tansella}, G.~{Jelic-Cizmek}, C.~{Bonvin} and R.~{Durrer}, \emph{{COFFE: a
  code for the full-sky relativistic galaxy correlation function}},
  \href{http://dx.doi.org/10.1088/1475-7516/2018/10/032}{\emph{\jcap}
  {\bfseries 2018} (Oct., 2018) 032},
  [\href{https://arxiv.org/abs/1806.11090}{{\ttfamily 1806.11090}}].

\bibitem{FOHOET13}
D.~{Foreman-Mackey}, D.~W. {Hogg}, D.~{Lang} and J.~{Goodman}, \emph{{emcee:
  The MCMC Hammer}}, \href{http://dx.doi.org/10.1086/670067}{\emph{\pasp}
  {\bfseries 125} (Mar., 2013) 306},
  [\href{https://arxiv.org/abs/1202.3665}{{\ttfamily 1202.3665}}].

\bibitem{MAFR00}
C.-P. {Ma} and J.~N. {Fry}, \emph{{Deriving the Nonlinear Cosmological Power
  Spectrum and Bispectrum from Analytic Dark Matter Halo Profiles and Mass
  Functions}}, {\emph{\apj} {\bfseries 543} (Nov., 2000) 503--513},
  [\href{https://arxiv.org/abs/astro-ph/0003343}{{\ttfamily
  astro-ph/0003343}}].

\bibitem{SMPEET03}
R.~E. {Smith}, J.~A. {Peacock}, A.~{Jenkins}, S.~D.~M. {White}, C.~S. {Frenk},
  F.~R. {Pearce} et~al., \emph{{Stable clustering, the halo model and
  non-linear cosmological power spectra}}, {\emph{\mnras} {\bfseries 341}
  (June, 2003) 1311--1332},
  [\href{https://arxiv.org/abs/arXiv:astro-ph/0207664}{{\ttfamily
  arXiv:astro-ph/0207664}}].

\bibitem{HEFRET21}
K.~{Heitmann}, N.~{Frontiere}, E.~{Rangel}, P.~{Larsen}, A.~{Pope}, I.~{Sultan}
  et~al., \emph{{The Last Journey. I. An Extreme-scale Simulation on the Mira
  Supercomputer}},
  \href{http://dx.doi.org/10.3847/1538-4365/abcc67}{\emph{\apjs} {\bfseries
  252} (Feb., 2021) 19}, [\href{https://arxiv.org/abs/2006.01697}{{\ttfamily
  2006.01697}}].

\bibitem{SZALA88}
A.~S. {Szalay}, \emph{{Constraints on the biasing of density fluctuations}},
  {\emph{\apj} {\bfseries 333} (Oct., 1988) 21--23}.

\bibitem{FRGA93}
J.~N. {Fry} and E.~{Gaztanaga}, \emph{{Biasing and hierarchical statistics in
  large-scale structure}}, \href{http://dx.doi.org/10.1086/173015}{\emph{\apj}
  {\bfseries 413} (Aug., 1993) 447--452},
  [\href{https://arxiv.org/abs/astro-ph/9302009}{{\ttfamily
  astro-ph/9302009}}].

\bibitem{FRY96}
J.~N. {Fry}, \emph{{The Evolution of Bias}},
  \href{http://dx.doi.org/10.1086/310006}{\emph{\apjl} {\bfseries 461} (Apr.,
  1996) L65+}.

\bibitem{SZMALA98}
A.~S. {Szalay}, T.~{Matsubara} and S.~D. {Landy}, \emph{{Redshift-Space
  Distortions of the Correlation Function in Wide-Angle Galaxy Surveys}},
  {\emph{\apjl} {\bfseries 498} (May, 1998) L1},
  [\href{https://arxiv.org/abs/arXiv:9712007}{{\ttfamily arXiv:9712007}}].

\bibitem{MATSU00}
T.~{Matsubara}, \emph{{The Gravitational Lensing in Redshift-Space Correlation
  Functions of Galaxies and Quasars}}, {\emph{\apjl} {\bfseries 537} (July,
  2000) L77--L80}, [\href{https://arxiv.org/abs/arXiv:0004392}{{\ttfamily
  arXiv:0004392}}].

\bibitem{SZAPU04}
I.~{Szapudi}, \emph{{Wide-Angle Redshift Distortions Revisited}}, {\emph{\apj}
  {\bfseries 614} (Oct., 2004) 51--55},
  [\href{https://arxiv.org/abs/arXiv:0404477}{{\ttfamily arXiv:0404477}}].

\bibitem{PASZ08}
P.~{P{\'a}pai} and I.~{Szapudi}, \emph{{Non-perturbative effects of geometry in
  wide-angle redshift distortions}}, {\emph{\mnras} {\bfseries 389} (Sept.,
  2008) 292--296}, [\href{https://arxiv.org/abs/arXiv:0802.2940}{{\ttfamily
  arXiv:0802.2940}}].

\bibitem{YOSE15}
J.~{Yoo} and U.~{Seljak}, \emph{{Wide-angle effects in future galaxy surveys}},
  \href{http://dx.doi.org/10.1093/mnras/stu2491}{\emph{\mnras} {\bfseries 447}
  (Feb., 2015) 1789--1805}, [\href{https://arxiv.org/abs/1308.1093}{{\ttfamily
  1308.1093}}].

\bibitem{PIMAET26}
G.~{Piccoli}, C.~{Magnoli}, J.~{Yoo} and D.~{Eisenstein}, \emph{{in
  preparation}}, {\emph{\jcap} {\bfseries 000} (2026) 10},
  [\href{https://arxiv.org/abs/arXiv:astro-ph/2600.00000}{{\ttfamily
  arXiv:astro-ph/2600.00000}}].

\end{thebibliography}\endgroup

\end{document}